\documentclass[aps,prl,twocolumn,preprintnumbers,groupedaddress,superscriptaddress,floatfix,tightenlines,reprint]{revtex4-1}
\usepackage{mathrsfs,natbib}
\usepackage{graphics,epsfig,color, graphicx}
\usepackage{verbatim}
\usepackage{bm}
\usepackage{enumerate}

\usepackage{graphicx,epsf,amssymb,bbm,amsbsy,amsfonts,amssymb,amsmath}
\usepackage{hyperref}

\newcommand{\eq}{\begin{equation}}
\newcommand{\eqe}{\end{equation}}

\def\tr{\text{tr}\,}

\newcommand{\eqa}{\begin{eqnarray}}
\newcommand{\eqae}{\end{eqnarray}}

\def\Z{{\mathbb Z}}
\def\R{{\mathbb R}}
\def\S{{\mathbb S}}

\usepackage{verbatim}   
\usepackage[all]{xy}
\usepackage{bm}  
\def\Dslash{{\rlap{\raise 1pt \hbox{$\>/$}}D}}
\def\Pslash{{\rlap{\raise  1pt \hbox{$\>/$}}\,\partial}}

\def\slashchar#1{\ensuremath{                               %
   \setbox0=\hbox{${}#1{}$}       
   \dimen0=\wd0                                 
   \setbox1=\hbox{/} \dimen1=\wd1               
   \ifdim\dimen0>\dimen1                        
      \rlap{\hbox to \dimen0{\hfil/\hfil}}      
      {}#1{}                                    
   \else                                        
      \rlap{\hbox to \dimen1{\hfil${}#1{}$\hfil}}   
   \fi}}

\hypersetup{
    colorlinks=true,       
    linkcolor=red,          
    citecolor=blue,        
    filecolor=magenta,      
    urlcolor=blue           
}

\begin{document}

\title{Emergent Dimensions and Branes from Large-N Confinement }

\preprint{INT-PUB-16-016}

\author{Aleksey Cherman}
\email{aleksey.cherman.physics@gmail.com}
\affiliation{Institute for Nuclear Theory, University of Washington, Seattle, WA 98105 USA}

\author{Erich Poppitz}
\email{poppitz@physics.utoronto.ca}
\affiliation{Department of Physics, University of Toronto, Toronto, ON M5S 1A7, Canada}

\begin{abstract}
$\mathcal{N}=1$ $SU(N)$ super-Yang-Mills theory on $\R^3\times \S^1$ is believed to have a smooth dependence on the circle size $L$. Making $L$ small  leads to calculable non-perturbative color confinement,  mass gap, and string tensions.  For finite $N$, the small-$L$ low-energy dynamics is described by a three-dimensional effective theory. The large-$N$ limit, however, reveals surprises: the infrared dual description is in terms of a theory with an emergent fourth dimension, curiously reminiscent of T-duality in string theory. Here, however, the emergent dimension is a lattice, with momenta  related to the $\S^1$-winding of the gauge field holonomy, which takes values in $\Z_N$. 
Furthermore, the low-energy description is given by a non-trivial gapless theory, with a space-like $z=2$ Lifshitz scale invariance and operators that pick up anomalous dimensions as $L$ is increased.   Supersymmetry-breaking deformations leave the long-distance theory scale-invariant, but change the Lifshitz scaling exponent  to $z=1$, and lead to an emergent Lorentz symmetry at small $L$. Adding 
   a small number of fundamental fermion fields leads to matter localized on three-dimensional branes in the emergent four-dimensional   theory. \end{abstract}


\maketitle

{\bf Introduction and summary. }
In this paper we explore the large-$N$ dynamics of pure $SU(N)$ $\mathcal{N}=1$ super-Yang-Mills (SYM) theory, a close cousin of Yang-Mills theory and QCD. When compactified on $\mathbb{R}^3 \times \S^1$ with periodic boundary conditions for fermions, 
this theory has the beautiful feature that color confinement, the mass gap, string tensions,  and chiral symmetry breaking can be studied analytically using semi-classical methods \cite{Davies:1999uw,Davies:2000nw,Unsal:2007jx,Poppitz:2012sw,*Poppitz:2012nz,Anber:2013doa,Anber:2014lba,Anber:2015kea}.  The calculable regime is obtained if $\eta =  L N \Lambda$ is small, where $\Lambda$ is the strong scale.  We thus study the large $N$ limit with $\eta\ll 1$ held fixed.

Usually, if a 4D theory with a mass gap lives on a circle, and the circle-dependence is smooth, the low-energy dynamics for small $L$ is described by a 3D effective field theory (EFT) with a gap.  
This is certainly the case for order-one values of $N$ for $\mathcal{N}=1$ SYM. At large $N$ and small $\eta$, however, we find two surprising features.   
First, the long-distance physics is described by a 4D EFT.  The fourth dimension emerges from the non-perturbative dynamics, via a mechanism different from e.g.~\cite{ArkaniHamed:2001nc,*Cheng:2001vd,*Steinacker:2014lma}.  Its  size $\tilde{L}$ is parametrically larger than both the circle size and the inverse strong scale: $\tilde{L} = L N^2/\eta^{3} = N/(\Lambda \eta^2)$, and its emergence resembles $T$-duality in string theory, at least superficially.  Second, the light glueball masses become parametrically separated by $N^2 \eta^{-3/2}$ from the scale of the 3D string tension,  and at large $N$ the 4D ``infrared dual" theory becomes a generically non-trivial gapless theory, with a spatial $z=2$ Lifshitz scaling symmetry.

We also study some supersymmetry-breaking deformations of $\mathcal{N}=1$ SYM, such as the addition of a gluino mass term or extra adjoint and fundamental fermion fields. The resulting theories are in the universality class of YM theory and QCD.  The surprising phenomena we found in SYM theory survive these deformations, with some interesting modifications.  A gluino mass term changes the Lifshitz parameter $z$ of the low-energy theory from $z=2$ to $z=1$.  Adding $N_f \ll N$ fundamental fermions leads to fields living on 3D branes in the emergent 4D bulk.

{\bf Phase structure and weak coupling.}
We now review some standard features of SYM theory. Readers interested in our main results may proceed to Eq.~\eqref{eq:sigma_potential}.  

It is believed that $\mathcal{N}=1$ SYM compactified on $\mathbb{R}^3 \times \S^1$ has a smooth dependence on the circle size $L$ so long as fermions have periodic boundary conditions on $\S^1$.  Indeed, at small $L$ the theory becomes weakly coupled, and one finds\cite{Davies:1999uw,Davies:2000nw,Unsal:2007jx} a Wilson loop  $\tr \Sigma = \tr \mathcal{P} e^{i\int_{\S^1}A_3}$ expectation value $\langle \Sigma \rangle\sim \mathrm{diag}(1, \omega,\ldots, \omega^{N-1})$, $\omega \equiv e^{2\pi i /N}$.  This means that $\langle \tr \Sigma^n \rangle = 0, \forall n<N$, signalling the preservation of the $\Z_N$ center symmetry and confinement even at small $L$, as expected from continuity in $L$.  It can also be checked that $\langle  {\psi} \psi \rangle \neq 0$ at small $L$, so that the discrete chiral symmetry is spontaneously broken, $\Z_{2N} \to \Z_2$, also as expected from continuity.

Thanks to the small-$L$ form of $\Sigma$, we find $\langle A_3\rangle \neq 0$, leading to a compact adjoint Higgs mechanism breaking the gauge group $SU(N) \to U(1)^{N-1}$.  So, at low energies compared to the lightest $W$-boson mass, $m_W \equiv 2\pi/(N L)$, the physics is described by an Abelian theory.  This is the reason why the small-$\eta$ physics is weakly coupled: all charged matter is at least as heavy as $m_W$, and the 't Hooft coupling $\lambda\equiv g^2 N$ stops running at the scale $m_W \gg \Lambda$, giving the weak-coupling condition $\eta \ll 1$.

The key point is that staying in the weak-coupling regime while making $N$ large requires an unusual scaling for the circle size, $L \sim \Lambda^{-1}/N$.  The physical reason such small values of $L$ are needed is that, for any $L\sim \mathcal{O}(N^0)$, $\mathcal{N}=1$ SYM theory \cite{Kovtun:2007py,Unsal:2008eg} enjoys large-$N$ volume independence \cite{EguchiKawaiOriginal,*Bhanot:1982sh,*GonzalezArroyo:1982ub,*GonzalezArroyo:1982hz}   and hence is strongly coupled.

{\bf Small-$L$ effective field theory.}
When $L$ is small and $\eta \ll 1$, the long-distance physics can be described by fields which carry zero momentum on $\S^1$.  Furthermore, the fields that do not get a mass from the adjoint Higgs mechanism  are the Cartan 3D gluons $F^{\mu \nu}_{k}$ ($k=1,\ldots,N-1$, $\mu,\nu=0,1,2$), Cartan gluinos $\psi_k^\alpha$ ($\alpha=1,2$), and Cartan scalars $\phi_k$, from fluctuations of $A_3$ (or $\Sigma$).  These fields have gauge-invariant representations such as 
\begin{align}
F^{\mu \nu}_{k} = \tfrac{1}{N}  \sum_{p=0}^{N-1}\omega^{-kp} \tr( \Sigma^p F_{\mu\nu})\,,
\label{eq:winding_Cartan}
\end{align}
so that the ``Cartan" index $k$ can be interpreted as the discrete Fourier transform of the winding number for the holonomy $\Sigma$.  So even though the fields $F^{\mu \nu}_{k}, \psi_k^{\alpha}, \phi_k$ carry no $\S^1$ momentum, they do carry information about the winding modes on $\S^1$.    
The fields $F^{\mu \nu}_{k}, \psi_k^{\alpha}, \phi_k$ all sit within a single supermultiplet, and are the fields that appear in the long-distance $\ell \gg m_W^{-1}$ EFT. 
 
 We will work with the Abelian dual representation of the gluons $F^{k}_{\mu \nu} 
=  g^2/(2\pi L) \epsilon_{\mu\nu\alpha} \partial^{\alpha} \sigma^{k}$, and take an $N$-component basis for the co-root vectors  $\vec{\alpha}^{*}_{i}$ of the $\mathfrak{su}(N)$ algebra satisfying $\vec{\alpha}^{*}_{i} \cdot \vec{e}_0 = 0, \forall i$ with  $\vec{e}_0\equiv(1,1,1, \ldots,1)$.  We package the fields into $N$-component $\vec{\sigma}, \vec{\phi}, \vec\psi$   vectors, whose e.g.  $\vec\sigma \cdot \vec e_0$ components are unphysical. 
For brevity, we focus on the $\vec\sigma$-field Lagrangian, omitting the $\vec \phi$ and $\vec \psi$  fields in Eqs.~\eqref{eq:sigma_potential}-\eqref{definem};  the full   superspace expressions are in \cite{Anber:2014lba}.

In perturbation theory, $\vec{\sigma}$ has a shift symmetry coming from the current conservation law due to the absence of color-magnetic monopoles.  Its action can be written as
 $S_{\sigma} = \int d^{3}x\, \lambda m_W (\partial_{\mu} \vec{\sigma})^2$.  
 
Non-perturbatively, the $\mathbb{R}^3 \times \S^1$-compactified SYM theory has $N$ types of BPS monopole-instanton configurations \cite{Lee:1997vp,Kraan:1998sn}, with $(\mathrm{magnetic},\mathrm{topological})$ charges $
\pm\left(\vec{\alpha}^{*}_i, 
1/N  \right)$, where $i$=$1$,...$N$,  and $\vec\alpha^*_N$ denotes the affine (lowest) co-root. Their  actions are $S_0 = S_{I}/N$, where $S_{I} = 8\pi^2/g^2$ is the action of the BPST instanton with charge $\pm (0,1)$.  In the absence of  a gaugino mass, each monopole-instanton carries two fermion zero modes.  Thus, even though monopole-instantons carry magnetic charge, they cannot directly generate a potential for $\vec{\sigma}$; instead, they generate a superpotential    \cite{Affleck:1982as}.  The superpotential can be used to deduce the form of the potential for $\vec{\sigma}$\cite{Seiberg:1996nz,Davies:1999uw,Davies:2000nw}.
One can also compute the $\vec{\sigma}$ potential directly, which reveals that it arises from non-BPS ``molecular" events called ``magnetic bions" \cite{Unsal:2007jx}, with charges $\pm( \vec{\alpha}^{*}_i - \vec{\alpha}^{*}_{i+1 ({\rm mod}  N)}, 0)$ and action $2S_0$.  

Taking all this into account, to leading non-trivial order in the semi-classical expansion, the action for $\vec{\sigma}$ is
\begin{align}
\!\!\!\!S_{\sigma} &= \int d^{3}x\, \bigg\{  \lambda m_W (\partial_{\mu} \vec{\sigma})^2  \nonumber\\
& + m_W^3 e^{-2S_0} \sum_{i=0}^{N-1}
\sin^2\left[\frac{1}{2}(\vec{\alpha}^{*}_{i ({\rm mod}  N)} - \vec{\alpha}^{*}_{i+1})\cdot \vec{\sigma}\right] \bigg\} ~.
\label{eq:sigma_potential}
\end{align}
   Equation~\eqref{eq:sigma_potential} contains a lot of physics \cite{Unsal:2007jx}. The fact that the $\vec{\sigma}$ potential is non-vanishing  implies that at finite $N$  the theory has a non-perturbative mass gap roughly of order $m_W e^{-S_0} \sim \Lambda\, \eta^2$; the latter form follows from the one-loop relation $8\pi^2/\lambda(\mu)  = 3 \log(\mu/\Lambda)$.
   
{\bf Emergent extra dimension.}  
We now discuss the quadratic actions for $\sigma, \phi, \lambda$ around one of the $N$ ground states more carefully, with a focus on the $N$-dependence.  The 3D EFT has $N$ chiral symmetry breaking vacua  $\langle \vec{\sigma} \rangle = \tfrac{2\pi k \vec{\rho}}{N}$, where $k = 0,1, \ldots, N-1$, $\vec{\rho} = \sum_{a=1}^{N-1} \vec{w}_a$ is the Weyl vector, and $\vec{w}_a$ are the fundamental weights, obeying $\vec{\alpha}_{a}^* \cdot \vec{w}_b = \delta_{ab}$.  Expanding around any given vacuum $\langle \vec{\sigma} \rangle$, we find the same quadratic action
\begin{align}
\!\!\!\!\!\!\!S_{\sigma} &= \int d^{3}x\, \sum_{i=1}^{N}\big\{ (\partial_{\mu} \tilde{\sigma}_i)^2 +  {M^2\over 16}(2\tilde{\sigma}_i -\tilde{\sigma}_{i-1} - \tilde{\sigma}_{i+1} )^2 \big\},
\label{eq:sigma_SYM_position_space}
\end{align}
where $\tilde{\sigma}$ represents fluctuations around $\langle \vec{\sigma} \rangle$, and we used the identity $(\vec{\alpha}^{*}_{i} - \vec{\alpha}^{*}_{i+1})\cdot \vec{\sigma}$ $=$ $2\sigma_i -\sigma_{i-1} - \sigma_{i+1}$, with all indices  $($mod $N)$.  So our results below apply to all $N$ chiral-symmetry-breaking vacua.  Omitting factors of order unity and non-exponential dependence on   $\lambda$, the mass scale $M$ is of order  
\begin{equation}\label{definem}
 a^{-1} := M \sim m_W e^{-S_0} \sim \Lambda\, \eta^2 ~
\end{equation}
where the  the ``lattice spacing"  $a$ = $1/M$ has been defined for future use. Parallel results hold for the quadratic actions for $\phi$ and $\psi$. The quadratic action, with all superpartners included, can be diagonalized by a discrete Fourier transform,
$
\{F_{p},S_{p},\Psi_{p} \}   = N^{-1/2} \, \sum_{n=1}^{N} \omega^{n p} \{\tilde{\phi}_n, \tilde{\sigma}_n,\tilde{\psi}_n\} 
$
leading to
\begin{align}
\!\!\!\!\!\!S_{\rm EFT} 
&= \int d^{3}x \,\sum_{p=1}^{N} \bigg\{  |\partial_{\mu} \Phi_p|^2+M^2 \sin^{4}\left(\frac{\pi p }{ N}\right)|\Phi_p|^2 \nonumber \\
&+ \bar{\Psi}_{p} \slashchar{\partial} \Psi_{p} +{ M \over 2 } \sin^{2}\left(\frac{\pi p }{N}\right) ({\Psi}_{N-p} \Psi_p + {\rm h.c.})  \bigg\},
\label{eq:sigma_SYM_momentum_space}
\end{align}
where $\Phi_{p} = F_{p} + i S_{p}$. Neglecting the fictitious $p=N$ modes, the EFT spectrum is 
\begin{align}
\label{spectrum1}
m_p = M  \sin^{2}\left(\frac{\pi p }{N}\right),\;\; p=1,\cdots, N-1.
\end{align}
This spectrum somewhat resembles the spectrum of some integrable 2D field theories\cite{Berg:1978zn,*Kurak:1978su,*Polyakov:1983tt,*Wiegmann:1984ec,*Fateev:1994ai}.   A physical interpretation of the index $p$ is given below in a discussion of T-duality.

The large-$N$ limit of the expressions above contains surprises. First, the mass gap \emph{vanishes} at large $N$, since the lightest mode has mass $m_1 \sim \Lambda \eta^2/N^2$.  Second, the states in Eq.~\eqref{spectrum1} with $p \sim N^0$ are quadratically-spaced, $m_p \sim m_1  p^2$.  Indeed, we recognize Eq.~\eqref{eq:sigma_SYM_position_space} as the  position-space action 
for a \emph{four-dimensional} field theory with one circular spatial direction, whose coordinate  we shall denote by `$y$', discretized on a lattice of spacing $a =1/M$,  
 and size $\tilde{L} = a N$.   Equation~\eqref{eq:sigma_SYM_momentum_space} is the momentum-space action with superpartners restored.  (Note in passing that there is no fermion doubling here.)

Thus, on large distances $\ell \gg a$ the low-energy effective theory turns into a continuum 4D theory for the fields $\Phi$ and $\Psi$.  The emergent dimension is compactified on a circle of size $\tilde{L} = a N$, and so looks non-compact on scales $\ell \sim \mathcal{O}(N^0) a$.  Indeed, $\tilde{L}$ is parametrically large compared to both the physical circle size $L$ and the inverse strong coupling scale $\Lambda^{-1}$:  $\tilde{L}/L \sim N^2 \eta^{-3}, \tilde{L} \Lambda = N \eta^{-2}$.  

To write the continuum $\tilde{L}$ $\gg$ $\ell$ $\gg a$ limit, one might naively try to scale lengths, times, and fields according to  the canonical 4D scaling dimensions, and replace e.g.~difference operators with derivatives as usual.   This gives continuum 4D fields $\Phi'$ and $\Psi'$ with kinetic terms of the form $a^2 |\partial_y^2 \Phi'|^2$ and $a {\Psi}' \partial_y^2  \Psi'$, which look technically irrelevant.
 However, this isotropic  assignment of scaling dimensions  is only natural in Lorentz-invariant 4D theories.  But we broke Lorentz invariance by compactifying SYM, and our emergent 4D theory is clearly not 4D Lorentz-invariant.   Instead, the low-energy theory enjoys  an \emph{anisotropic} ``spatial Lifshitz" scale invariance\cite{Hornreich:1975zz}
\begin{align}
\label{scalingz}
\begin{matrix}
~~~~x_{0,1,2} \to \Omega \; x_{0,1,2}, & y \to \Omega^{1/z} \; y\,\\
~~~~\Phi \to \Omega^{-(1+1/z)/2} \; \Phi, &~~~~\Psi \to \Omega^{-(2+1/z)/2} \;\Psi \, ~,
\end{matrix}
\end{align}
with $z=2$.  With appropriately rescaled coordinates and fields, the Lifshitz-scale-invariant continuum limit can thus be written as 
\begin{align} 
\!\!\!\!\!\!\!S  
 =\!\!\int\!\! d^{3}x\,d y  \{  |\partial_{\mu} \Phi|^2 \!+\! |\partial_y^2 \Phi|^2 \!+\! \bar{\Psi}  \slashchar{\partial}\Psi \! +\! \tfrac{1}{2}( \Psi \partial_y^2   \Psi \!+\! {\rm h.c.})\}. 
\label{eq:Lifshitz_action}
\end{align}
This is one of our key results. Higher order terms from the expansion of Eq.~\eqref{eq:sigma_potential} are irrelevant under the $z=2$ Lifshitz scaling.  Thus the long-distance large-$N$ theory is free to leading order in $\eta \ll 1$. The gapless continuum theory of Eq.~\eqref{eq:Lifshitz_action} describes the physics on length scales $\ell \sim \mathcal{O}(N^0)$ satisfying $\ell \gg a$ when $\eta \ll 1$. 

{\bf Symmetries and corrections.}  Equation~\eqref{eq:sigma_potential}, which led to Equation~\eqref{eq:Lifshitz_action}, contains only the leading terms in the $\eta \ll 1$ semi-classical expansion.  We now argue that higher-order corrections  cannot produce a large-$N$ mass gap. The possibility of describing the infrared (IR) fixed point in terms of a scale-invariant local free-field theory, as in Eq.~\eqref{eq:Lifshitz_action}, turns out to be tied to the symmetries of the long distance theory, along with the weak coupling limit $\eta \ll 1$.  Our central observations also apply to non-supersymmetric theories.

First, consider the mass gap.  Due to charge quantization,  $\vec{\sigma}$ is a compact variable living  in the unit cell  generated by the $N-1$ fundamental weights  $\vec{w}_{k}$ of  $\mathfrak{su}(N)$.  Thus the effective action must be periodic under $\vec{\sigma} \to \vec{\sigma} + 2\pi \, \vec w_k, \, \forall k$, and the $\vec{\sigma}$ potential must be a function of $e^{i \vec{\alpha}^{*}_i \cdot \vec{\sigma}}$, since $\vec{\alpha}^{*}_i \cdot \vec{w}_j = \delta_{ij}$. 
Further, the action must be invariant under  the $\mathbb{Z}_N$ center symmetry, acting  \cite{Anber:2015wha} as $ \sigma_{i} \to \sigma_{i+1(\textrm{mod } N)}$, or  as $\vec{\alpha}^{*}_i \cdot \vec{\sigma}  \to \vec{\alpha}^{*}_{i+1 (\textrm{mod } N)} \cdot \vec{\sigma}$.  
Finally, the $\Z_N$ subgroup of the discrete chiral symmetry acts $e^{i \vec\alpha_i^* \cdot \vec\sigma} \to e^{i {2 \pi \over N}} e^{i \vec\alpha_i^* \cdot \vec\sigma}$.  These chiral shifts become continuous at $N=\infty$, but domain wall tensions stay large\cite{Dvali:1996xe,*Kovner:1997ca} and there is no light $\eta'$ mode.
 
Remarkably, the $\vec\sigma$-periodicity condition, together with $\vec\alpha_i^* \cdot \vec\sigma = \sigma_i  - \sigma_{i+1}$, implies that corrections to the $\vec\sigma$ potential can only ever produce terms which look like discretized derivatives in the $y$-coordinate.   The role of large $N$ is to allow a continuum limit in which the theory acts as if it lives on a large circle of size $\tilde{L} \sim N$.  Thus the lowest emergent KK momentum goes to zero with $N$.  At the same time, a mass term  $\int d^3{x}  \sum_{i=1}^{N} \sigma_i^2 \sim \int d^3 x \,dy\, \Phi^{\dag} \Phi$, can never be generated either perturbatively or non-perturbatively in $\eta$, or perturbatively or non-perturbatively in the $1/N$ expansion.  The microscopic reason for this is that such a mass term is forbidden by the discrete gauge symmetry which imposes compactness of the $\vec{\sigma}$ variable.    So, in the domain smoothly connected to $\eta \to 0$, that is, for some range $\eta \in (0, \eta_c > 0)$, our results imply that no mass gap can be generated at large $N$.  It is important to note that the arguments above are completely independent of supersymmetry:  they hold so long as center symmetry is preserved. 
 
Next, consider the scale invariance and locality of the action Eq.~\eqref{eq:Lifshitz_action}.   In a scale-invariant theory, the availability of a local Lagrangian description is tied to whether the theory is free.  We now show that the IR fixed point becomes free as $\eta \to 0$, but is in general non-trivial.

First, note that no local (in $y$) Lifshitz-breaking interactions, such as $(\partial_y \Phi)^2$, can be induced either perturbatively or nonperturbatively. This is due to masslessness of the gauginos and the consequent discrete $\Z_N$ chiral symmetry, which forbids monopole-instantons from directly generating a bosonic potential.   (The discrete chiral symmetry, along with Lifshitz scaling, is broken when a SUSY breaking mass is added, see further discussion below.)  One may worry that the chiral and center symmetries would permit nonlocal in $y$ terms like Re $\sum_k e^{i \vec\alpha_k^* \cdot \vec\sigma} e^{ - i \vec\alpha_{k + N/2}^* \cdot \vec\sigma}$, but such terms do not arise at small $\eta$. In SYM,  potential terms come from the superpotential, determined by symmetries and holomorphy to be  \begin{align}
  \label{super1}
  W = \Lambda^2  \; \eta \; e^{i {2 \pi k \over N}} \; \sum_{l=1}^{N} e^{\vec\alpha^{*}_{l} \cdot \vec{X}} \, ,
  \end{align}
where $\vec{X}$ is a chiral superfield whose lowest component is $\vec{\phi}+i \vec{\sigma}$. (This is the  superpotential from \cite{Davies:1999uw,Davies:2000nw}, but with $\vec{X}$ defined with its expectation value in the $k^{\rm th}$ vacuum subtracted, so that $(\partial W/\partial \vec{X})\vert_{\vec{X}=0}=0$.) The superpotential Eq.~\eqref{super1} is not renormalized and, with the canonical K{\"a}hler potential $K\sim\vec{X}^\dagger \cdot \vec{X}$, gives rise to Eq.~\eqref{eq:sigma_potential}. Expanding Eq.~\eqref{super1} to quadratic order in the fields allows us to cast $W$ as an integral over the extra dimension
 \begin{align}
\!\!\!\! \!\!\! W \sim \Lambda^2\, \eta\, e^{i {2 \pi k \over N}} \sum_{i=1}^{N} (\nabla_+ X^i)^2 \rightarrow {\Lambda e^{i {2 \pi k \over N}} \over \eta}\! \int \!dy \; (\partial_y X)^2 \,,
  \end{align}
where $\nabla_+ X^i = X^{i+1} - X^i$ and we took the same continuum limit that led to Eq.~\eqref{eq:Lifshitz_action}.

As usual, none of the symmetries forbid the generation of a non-trivial K\" ahler potential, which has the effect of changing the 3D kinetic terms for $\vec{\sigma}$ from $\sum_{i}(\partial_{\mu} \sigma_{i})^2  \to \sum_{i,j} G^{-1}_{ij}\partial_{\mu} \sigma^{i} \partial^{\mu} \sigma^{j}$. The inverse K\" ahler metric $G_{ij}$ is determined by the moduli space metric of the electric theory after a linear-chiral duality transformation \cite{Anber:2014lba}; at the center symmetric point, it is a function of  $\eta$ and $N$ only. The leading-order correction to the flat K{\"a}hler metric comes from threshold corrections due to loops of e.g.~massive $W$-bosons. The explicit calculation \cite{Anber:2014lba} shows that the large $N$ form of $G_{ij}$ is 
\begin{align}
G_{ij} = \delta_{ij} \frac{1}{c\, \lambda}+ {1 - \delta_{ij} \over | i - j|} \, ,
\end{align}
where $c = 3/16\pi^2$ and $\lambda$ is the 't Hooft coupling at the scale $m_W = 2\pi/LN$.
The continuum-limit two-point function of the field $\Phi$ then becomes
\begin{align}
\!\!\!\!\!\!\!\!\!\int d^4x \, e^{i p_{M} x^{M}}\! \langle \Phi^{\dag}(x_{M}) \Phi(0)\rangle \! \sim\! \frac{1}{p_y^{ 2c  \lambda } \left(p_{\mu}^2 +p_{y}^{4\left[1- c  \lambda \right]}\right)}\,,
\label{eq:anomalous_scaling}
\end{align}
where $M = 0,1,2,y$.  
This amounts to an $L$-dependent anomalous dimension for the $\Phi$ field, because $\lambda \sim 1/\log(1/\eta)$.  So $\mathcal{N}=1$ SYM on $\mathbb{R} \times \S^1$ flows to a four-dimensional non-trivial scale-invariant fixed point in the IR at large $N$ within the regime smoothly connected to $\eta \sim 0$.   The IR fixed point develops a local free-field representation as $\eta \to 0$, where the scale symmetry is the $z=2$ Lifshitz invariance of Eq.~\eqref{eq:Lifshitz_action}.  
 
{\bf String theory versus gauge theory.} 
Confining gauge theories with adjoint matter are believed to be weakly-coupled closed string theories with $g_s \sim 1/N$ in the usual large $N$ limit. In contrast, here we study a large $N$ limit with small $\eta$, where the gauge theory itself is weakly coupled.  Thus any dual string description it may have is bound to be quite distinct from e.g. the standard gauge-gravity duality\cite{Aharony:1999ti}. Nevertheless, our theory has several tantalizing features that make a string theory connection worthy of a further look.

String theory on a small circle $L$ is usually equivalent to another string theory on an `emergent' large circle $\tilde{L}$, with  $L \tilde{L} \sim \alpha'$, where $\alpha'$ is the inverse string tension.  Here, we study  a gauge theory on a tiny circle $L$, where it continues to confine, and find that at large $N$ it looks like another QFT on a large circle $\tilde{L}$.  Could this be some shadow of T-duality?  If so, at least two features ought to be present.  First, the emergent Kaluza-Klein (KK) momentum should have an interpretation as a winding number.  Second, it should be the case that $L \tilde{L} \sim \alpha'$ where $1/\alpha'$ is the confining string tension.  

Indeed, the emergent KK momentum $p$ in Eq.~\eqref{spectrum1} is in one-to-one correspondence to the winding number $p$ of the holonomy $\Sigma$ from Eq.~\eqref{eq:winding_Cartan}. To see this, note that the discrete Fourier transform leading to Eq.~\eqref{eq:sigma_SYM_momentum_space} is the inverse of the transform in Eq.~\eqref{eq:winding_Cartan}.  The emergent KK momentum is discretized, in contrast to the T-duality story for fundamental strings. Some insight into why this happens comes from recalling that fundamental strings  have winding numbers $w$ taking values in $\mathbb{Z}$, while gauge field holonomy winding numbers take values in $\mathbb{Z}_{N}$.  This is because $\langle \tfrac{1}{N} \tr \Sigma^N\rangle = 1$ regardless of the phase of the theory, since $N$ quarks can make a colorless state, a baryon, and widely separated baryons and anti-baryons do not interact via a color flux tube.  

Despite the pleasing resonance with T-duality intuition explained above, we do not know how to understand the fact that the emergent dimension is a Lifshitz one from this perspective.  The issue is that in SYM we find the scaling $(\mathrm{energy} \sim w^2)$, while all the weakly-coupled string models we are aware of give the scaling  $(\mathrm{energy} \sim w)$.   However, as explained in the next section, once SUSY is broken by e.g. a gaugino mass term, we find an emergent 4D Lorentz invariance at long distances, corresponding to the naively expected stringy scaling of $(\mathrm{energy} \sim w)$.

We now comment on the relation of the size of the emergent dimension to the string tension.  Even at small $\eta$,  the string tension in SYM for generic $N$ is only known from the  estimates of  \cite{Unsal:2007jx,Anber:2014lba,Anber:2015kea}.  These estimates naively suggest two different string tensions in SYM on $\mathbb{R}^3 \times \S^1$.  One of them, $1/\alpha'_{3D} \sim \Lambda^2 \eta$, is for strings stretched along $\mathbb{R}^3$, and the other, $1/\alpha'_{\S^1} \sim \Lambda^2 \eta/N$, is for strings winding around the $\S^1$.  The $\mathbb{R}^3$ strings can be made arbitrarily long compared to their width, so the definition of their string tension is unambiguous.   Amusingly, we indeed see that $L \tilde{L} \sim \alpha'_{3D}$.   The proper definition of the string tension for strings winding $\S^1$ is less obvious, because they cannot be made arbitrarily long. Refs.~\cite{Unsal:2007jx,Anber:2014lba,Anber:2015kea} defined $\alpha_{\S^1}$ from the scale of the exponential in the Polyakov-loop correlator, but how seriously one can take this definition is not clear.  Taking the definition at face value  we get $L \tilde{L} \sim N\alpha'_{\S^1}$, which is not what one would expect from a T-duality picture.  

To summarize the story, it is not yet obvious that there is a sharp connection between the emergent dimension phenomenon we have uncovered in gauge theory to T-duality.  But there are enough parallels that make the issue is worthy of  further study.  

Another property which is interesting to observe is that when $\eta \ll 1$ there is a large parametric separation
\begin{align}\label{separation}
\frac{\sqrt{1/\alpha'_{3D}}}{m_1}  \sim N^2\; \eta^{-3/2}
\end{align}
 between the light ``glueball" masses and the 3D string tension.  Parametric scale separations between glueball and string tensions scales are known to occur for gauge theories with supergravity duals in AdS/CFT \cite{Aharony:1999ti}, where the separation is governed by the strong 't Hooft coupling\footnote{We thank L.~Yaffe for emphasizing this to us.}. It is intriguing  to see the scale separation in a QCD-like theory, where it is controlled by the \emph{weak} 't Hooft coupling --- the $\eta^{-3/2}$ factor in Eq.~\eqref{separation} --- along with an $N$-dependent factor absent in holographic constructions.  As will be clear below, this scale separation also persists after SUSY-breaking deformations.

{\bf SUSY breaking and emergent Lorentz symmetry.}~We now discuss the effects of supersymmetry breaking. Turning on a gaugino mass term $ {N m_{\psi} \over \lambda}\;  \tr {\psi}\psi$ breaks SUSY and puts the theory in the universality class of pure YM theory. As shown in \cite{Poppitz:2012sw,*Poppitz:2012nz}, for $m_\psi \lesssim m_1$, see Eq.~\eqref{spectrum1}, the spectra of $\sigma$, $\phi$, and $\psi$ become split from each other due to the breaking of SUSY (for example, the $\phi$ fields become lighter, since increasing $m_\psi$ at fixed $\eta$ eventually leads to a first-order center-symmetry breaking phase transition). 

To get a confining regime for a wider range of $m_{\psi}$, we add an extra periodic adjoint fermion $\chi$, with mass $m_{\chi}$, to the SYM theory with gaugino mass $m_\psi$. This stabilizes center symmetry at small $\eta$ even if $m_{\chi} \sim m_{\psi} \sim m_W$\cite{Unsal:2008ch,Myers:2009df,Unsal:2010qh}. 
The fermion fields and $\vec{\phi}$ decouple at long distances, and the long-distance EFT only contains $\sigma^{i}$.
The $\sigma^{i}$ potential is now different because monopole-instantons can directly contribute to it, because there are no fermion zero modes with massive fermions.   This gives \cite{Unsal:2008ch} the long distance effective action  
\begin{align}
S_{\sigma,\rm  dYM} &= \int d^{3}x\, \bigg\{  \lambda m_W (\partial_{\mu} \vec{\sigma})^2  \nonumber\\
& +m_W^2 m_{\psi} e^{-S_0} \sum_{i=0}^{N-1}
\sin^2\left[\frac{1}{2}\vec{\alpha}^{*}_{i}\cdot \vec{\sigma}\right] +\cdots\bigg\}  \,
\label{eq:YM_sigma_potential}
\end{align}
where $\cdots$ represents higher order semiclassical contributions.  The quadratic action now includes $(\sigma_{i} - \sigma_{i+1})^2$, and the spectrum takes the form $m_{p}^2 \sim M^{*}{}^2 \sin^2(\pi p/N)$, where $M \sim \sqrt{M m_{\psi}} =: a_{*}^{-1}$.  

Thus, on large distances $\ell$ satisfying $a_{*} \ll \ell \ll \tilde{L}_{*}$, where  $\tilde{L}_{*} = N a_{*}$, the EFT is now a single gapless scalar field $S$ propagating in four dimensions with a kinetic term of the form $(\partial_{y} S)^2$ rather than $(a_*)^2 (\partial_{y}^2 S)^2$.   The  natural scale invariance is the conventional $z=1$ isotropic 4D scaling $x_{0,1,2} \to \Omega x_{0,1,2}, y \to \Omega y_{0,1,2}$. So, at long distances, with broken SUSY the large $N$ IR theory becomes a gapless Lorentz-invariant 4D scalar field for $\eta \ll 1$. Of course, just as in the SUSY case, we expect W-boson loops to generate small (when $\eta \ll 1$) non-Lorentz-invariant anomalous dimensions in the two-point function of $S$.  In any case, we see that even when SUSY is broken, at large $N$ the long-distance EFT is still a 4D scale-invariant QFT.

{\bf Fundamental fermions and branes.} 
We now consider adding $N_f$ fundamental fermions to $\mathcal{N}=1$ SYM.   To keep center symmetry at small $\eta$ after this deformation, we again need to add an extra periodic massive adjoint fermion $\chi$ with $m_{\chi} \lesssim m_W$ \cite{Poppitz:2012sw,*Poppitz:2012nz,Unsal:2010qh}.  The  $\phi_{k}$ modes then get masses $m_{\phi_k} \sim m_W$ and decouple in the IR.

Suppose $N_f=1$ and call the fundamental fermion $q$, with boundary condition $q(x_3+L) = e^{i\alpha} q(x_3)$.  In the center-symmetric background with $\alpha=0$, only one color component of $q$ remains massless, say $q_1$.  Index theorems\cite{Nye:2000eg,Poppitz:2008hr} imply that $q_1$ couples to a single monopole-instanton, so the $q$-part of the low-energy EFT is, schematically,
\begin{align}
\!\!\!\!\!S_{q}  = L\int d^3x \bigg\{  \bar{q}_1\big(\slashchar{D} 
+\frac{i\alpha}{L} \big)q_1  
 + m_W^{-2} e^{-S_0}  e^{i\vec{\alpha}^{*}_{1} \cdot \vec{\sigma}}  {\psi}_1  \psi_2  \bar{q}_{1  } q_{1  } \bigg\}.
 \label{eq:braneworld}
\end{align} 
Here $D_{\mu} = \partial_{\mu} + i g A_{1,\mu}$ since the quark has color-electric charge, and $\alpha$ plays the role of a real mass.  This 3D QED theory coupled to a fermion $q$ remains weakly coupled so long as $ \lambda/N \lesssim \alpha \lesssim 1/N$.

\begin{figure}[t]
\centering
\includegraphics[width=0.5\textwidth]{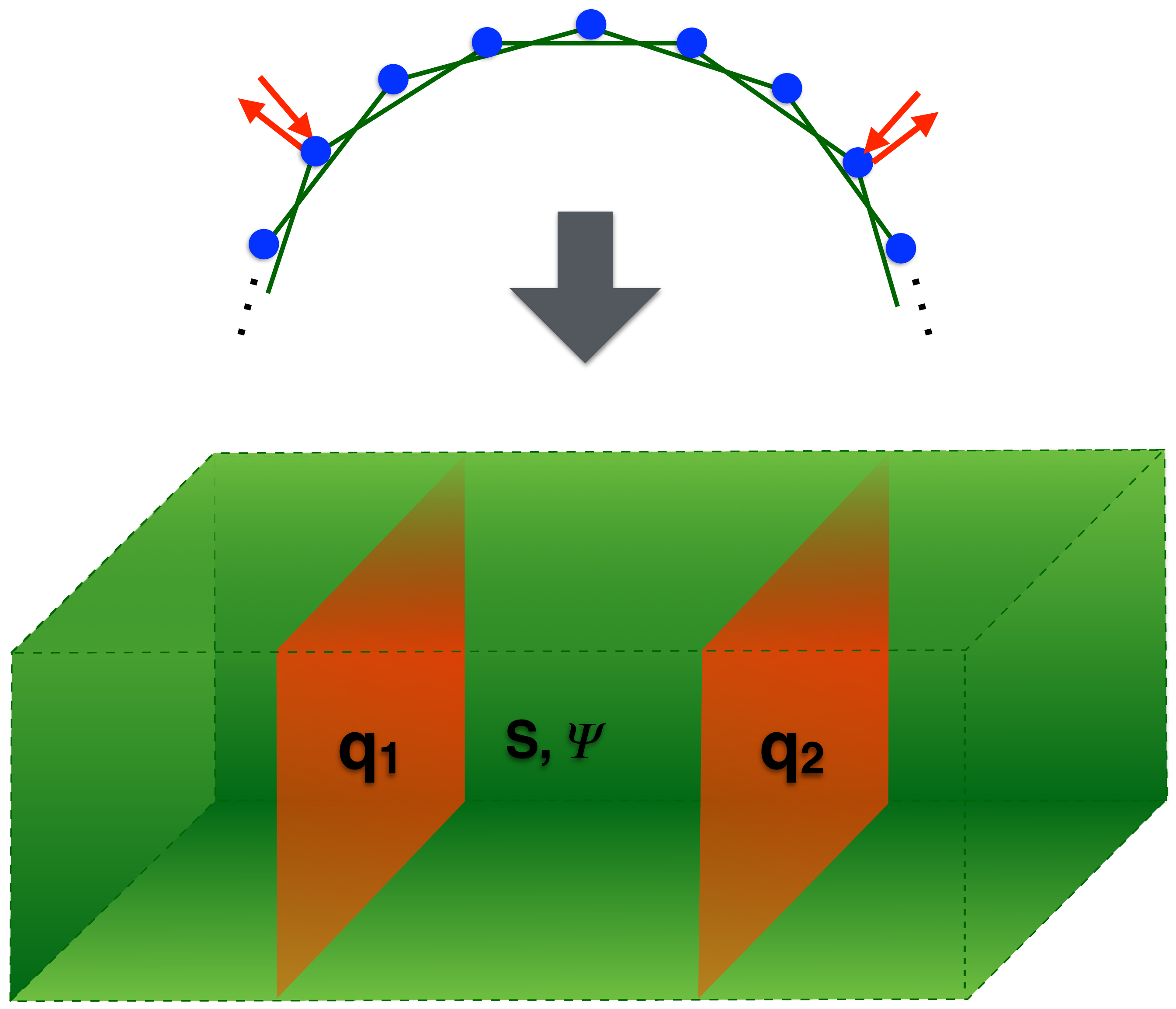}
\caption{(Color online.) Emergence of branes from large $N$ confinement at small $L$ with $N_f=2$.  At the top, the blue dots are $\sigma_{i}$ fields,
green lines indicate their interactions, and red arrows indicate fundamental fermion zero mode couplings.  On the bottom,  the 4D bulk, where the fields $S$ and $\Psi$ propagate, is in green, while two 3D branes, each carrying a flavor of the fundamental fermions, are in red. }
\label{fig:branes}
\end{figure}

The main point of Eq.~\eqref{eq:braneworld} is that fundamental fermions do \emph{not} propagate into the emergent fourth dimension.  
From the point of view of the low-energy EFT, adding $N_f$ fundamental fermions gives rise to matter localized on $N_f$ three-dimensional branes in the emergent four-dimensional bulk.  Flavor-dependent boundary conditions can be used to separate the branes by dialling which monopole-instantons pick up the fundamental quark zero modes, as explained in \cite{Cherman:2016hcd}, and the flavor branes become coincident when the boundary conditions are identical.  
We illustrate the situation in Fig.~\ref{fig:branes}.

{\bf Outlook and implications.}
We have explored the large $N$ behavior of a systematically calculable regime of confining gauge theories, defined by $\eta = N L \Lambda \ll 1$, and found several surprises concerning $\mathcal{N}=1$ SYM theory and some of its non-supersymmetric deformations.

First, the low-energy EFT description of SYM theory on $\mathbb{R}^3 \times \mathbb{S}^1$ becomes four-dimensional at large $N$, even though the original QFT is in the small-circle limit.  The fourth dimension is invisible in perturbation theory, and is an emergent large-$N$ consequence of the non-perturbative confining dynamics.  The second surprise is that the field content of this EFT is gapless at large $N$.  Indeed, we find that within the domain of validity of  the semiclassical small-$\eta$ expansion, symmetry considerations imply that at large $N$, a mass gap for the lightest fields $\sigma_i$ cannot be generated \emph{either perturbatively or non-perturbatively}.  So the large $N$ theory is gapless for $\eta \in (0, \eta_c)$ with some $\eta_c>0$. In particular, for small $\eta$ and large $N$, we find that the long-distance dynamics of $\mathcal{N}=1$ SYM is described by a non-trivial four-dimensional scale-invariant field theory with $z=2$ Lifshitz scale invariance.  This IR fixed point gets a free-field description as $\eta \to 0$, but in general the fields have non-vanishing anomalous dimensions at the fixed point.  The scale-invariant long-distance description is valid for distances $\ell$ which are big compared to $\eta^{-2} \Lambda^{-1}$ and small compared to the scale $N\eta^{-2} \Lambda^{-1}$.   These surprises motivate further careful lattice studies  of  SYM theory, along the lines of e.g. \cite{Fleming:2000fa,*Giedt:2008xm,*Endres:2009yp,*Demmouche:2010sf,*Bergner:2012rv,*Bergner:2013nwa,*Bergner:2015adz}.

For comparison, we note that an Abelian large-$N$ limit of softly-broken Seiberg-Witten (SW) theory similar to one considered here was studied in  \cite{Douglas:1995nw}.  In the infinite-$N$ limit, \cite{Douglas:1995nw} also found a gapless spectrum, but with two major differences compared to our large-$N$ results for SYM on $\R^3 \times \S^1$.  First, in SW theory, the spectrum has no known local extra-dimensional interpretation, due to the lack of an unbroken zero-form \cite{Gaiotto:2014kfa} center symmetry. Second, the string tensions and mass gap vanish simultaneously in SW theory, while in SYM the string tensions stay finite at large $N$. 

We now comment on the obvious questions raised by our results:  
\begin{enumerate} 
\item[{\bf Q1.}]{What are the implications and origin of the extra dimension?} 
\item[{\bf Q2.}]{Why is the long-distance theory is gapless?}
\item[{\bf Q3.}]{Is $\eta_c$ finite, for instance $\eta_c \sim 1$, or is it infinite?}
\end{enumerate} 

Concerning {\bf Q1}, the emergence of the extra dimension clearly has consequences for e.g.~the thermodynamics and transport properties at large $N$.  In SYM, for instance, the emergence of non-trivial Lifshitz scaling means that the thermodynamic and transport properties will be ``anomalous", serving as an illustration of some of the ideas in \cite{Hartnoll:2015sea,*Karch:2015pha} in a simple context.  In particular, the thermodynamics and transport will not be that of a typical 3D QFT even in the small $S^1$ limit.

The  conceptual origin of the extra dimension is not  yet clear. There are some striking, but at present superficial, parallels between our story and T-duality in string theory.   From another perspective, in some ways our extra dimension seems like a small-$\eta$ shadow of the complete large $N$ volume dependence that emerges for $\eta \gg 1$ via a working version of Eguchi-Kawai reduction \cite{Kovtun:2007py,EguchiKawaiOriginal,Bhanot:1982sh,GonzalezArroyo:1982ub,GonzalezArroyo:1982hz}. But in other ways the story appears to be very different.   

To appreciate this, first recall that in the absence of center-symmetry-breaking phase transitions (which are not expected in e.g. $\mathcal{N}=1$ SYM theory on a spatial circle), large $N$ volume independence is expected to set in smoothly as $\eta$ becomes large, and does not just emerge suddenly at $\eta = \infty$.  This is because, as shown in  \cite{Gross:1982at}, planar perturbation theory in a theory compactified on e.g. $\S^1$ of size $L$ in a $\mathbb{Z}_N$ center-symmetric background is identical to the planar perturbation theory of a theory compactified on a circle of size $NL$.   So, as long as the large $L \Lambda$ limit is smooth, and the large $L \Lambda$ and large $N$ limits commute, which is certainly expected, volume independence should set in smoothly as $\eta$ is increased.  In particular, at large $N$ and large $\eta$, the confining dynamics of a theory compactified on $\mathbb{R}^3 \times \S^1$ conspires to make it behave like a theory on $\mathbb{R}^4$.  A modern discussion of such phenomena in the context of lattice gauge theory on $T^4$ can be found \cite{Perez:2014sqa}.   

But, in a sharp contrast to this 1980s story, in our small-$\eta$ situation, the fourth dimension emerges out of the non-perturbative dynamics, and is \emph{invisible} to any order in perturbation theory.  Volume-independence analogies also give no insight into why the long-distance large $N$ theory should be gapless.  Indeed, this motivates turning to {\bf Q2}.

At large $N$, we have found that the gap is zero.  The symmetries of the microscopic theory forbid a mass term in the long-distance EFT, so a large $N$ mass gap cannot arise perturbatively or non-perturbatively.  But what is the conceptual origin of the gaplessness of the IR theory? On general grounds, a non-empty IR fixed point is natural if either (a) there are no relevant operators which could trigger a further RG flow or (b) there are relevant operators, but they are charged under a global symmetry of the fixed-point theory, so that one can naturally set them to zero.   Our theories certainly have relevant operators, $\int d^3{x}\, dy\, \Phi^{\dag}\Phi$, or $\int d^3{x}\, dy\, S^2$ in the non-supersymmetric cases,  but they are never generated by the microscopic dynamics.  It is not obvious to us how to interpret this in terms of the symmetries of the IR fixed point.  Understanding this better is important for theoretical and phenomenological reasons, since there are very few known ways to get naturally gapless IR scalars out of interacting QFTs.

Finding a long-distance theory built from scalar fields with irrelevant derivative interactions makes it tempting to wonder if the gaplessness is due to the spontaneous breaking of some continuous global symmetry. Such a symmetry would have to emerge only at large $N$, since the finite-$N$ gauge theory certainly has a gap.  (One candidate might have been the $\Z_N \rightarrow U(1)$ one-form \cite{Gaiotto:2014kfa} electric center symmetry, but it is not spontaneously broken in our theory.)  Whatever this large $N$ symmetry might be, it must survive even without supersymmetry given our results on SUSY-breaking deformations.  The possibility that large-$N$ confining theories might have rich emergent symmetries has recently been emphasized in \cite{Basar:2013sza,*Basar:2014hda,*Basar:2014jua,*Basar:2015asd,*Basar:2015xda,*Cherman:2015rpc}.

Finally turning to {\bf Q3}, we note that either a finite or an infinite $\eta_c$ would be remarkable.   A finite $\eta_c$ would mean that mass gap of large $N$ confining theories vanishes at some circle size.  Using the gap as an order parameter  would then imply a large-$N$ phase transition in the circle-size dependence of $\mathcal{N}=1$ SYM theory.  This would be striking, because then at large $N$ a SUSY-preserving compactification would not yield a smooth dependence on the circle size, contrary to expectations.    

On the other hand, if $\eta_c$ is infinite, we see two possible interpretations, either of which would be striking.  It could be that in fact $\eta_c \sim N$.  This would imply a non-commutativity of the large $L\Lambda$ limit with large $N$ limit for observables like the mass gap, as explained in our comments on volume independence. We know of no reason to expect this.  If the large $L\Lambda$ and large $N$ limits commute, and if $\eta_c$ is infinite, then then the gapless sector will survive into the volume-independent large-circle regime.  This would imply that confining gauge theories with a gap at finite $N$ can develop a gapless sector at large $N$, even on $\mathbb{R}^4$! 

This last option is so surprising that we add a few comments.  The reason that $\mathcal{N}=1$ SYM is believed to have a mass gap on $\mathbb{R}^4$ is that it has discrete chiral symmetry breaking and finite domain wall tensions, and is expected to have a finite confining string tension, for instance due to existing lattice studies\cite{Fleming:2000fa,*Giedt:2008xm,*Endres:2009yp,*Demmouche:2010sf,*Bergner:2012rv,*Bergner:2013nwa,*Bergner:2015adz}.  It is widely expected, on heuristic grounds, that these features have to be associated with a theory with a finite mass gap.  

Our calculations serve as an explicit counterexample to this expectation.  At large $N$, the small-$\eta$ regime of SYM  has discrete chiral symmetry breaking, finite domain wall tensions, and finite string tensions, and yet it has a vanishing mass gap.  The full QFT is of course not scale invariant, and so the theory will also have massive bound states in its spectrum.  There is no obvious conflict with the existing lattice studies\cite{Fleming:2000fa,*Giedt:2008xm,*Endres:2009yp,*Demmouche:2010sf,*Bergner:2012rv,*Bergner:2013nwa,*Bergner:2015adz}.  The fact that our results survive SUSY breaking raises the tantalizing possibility that these surprising phenomena could be present in theories like YM theory and QCD at large $N$.  Clearly, there is still a lot left to understand about large-$N$ confining dynamics.

{\bf Acknowledgments.}  We are grateful to T.~D.~Cohen, D. Gross, K.~Grosvenor, D.~B. Kaplan, A.~Karch, D.~McGady, M.~Rangamani, T.~Sulejmanpasic, M.~Wagman, and especially M.~\"Unsal, E.~Witten, L.~Yaffe, and an anonymous referee for helpful discussions.  A.~C. thanks the University of Toronto for hospitality at the initial stages of this project, and we thank the Fine Theoretical Physics Institute at the University of Minnesota for hospitality near its conclusion.  This work 
was supported in part by the U. S. Department of Energy under the grant DE-FG02-00ER-41132 (A.~C.) and an NSERC Discovery Grant (E.~P.)

\bibliographystyle{apsrev4-1}
\bibliography{small_circle} 
\end{document}